\documentclass[prx,twocolumn,showpacs,superscriptaddress]{revtex4}

\usepackage{graphicx}
\usepackage{mathrsfs}
\usepackage{dcolumn}
\usepackage{bm}
\usepackage{amsmath}
\usepackage{amsfonts}
\usepackage{color}

\begin{document}

\title{Dirac semimetal in type IV magnetic space groups}

\author{Guiyuan Hua}
\thanks{These authors contributed equally to this work.}
\affiliation{Wuhan National High Magnetic Field Center and School of Physics, Huazhong University of Science and Technology, Wuhan 430074, China}
\author{Simin Nie}
\thanks{These authors contributed equally to this work.}
\affiliation{Department of Materials Science and Engineering, Stanford University, Stanford, California 94305, USA}
\author{Zhida Song}
 \email{song.zhida@iphy.ac.cn}
\affiliation{Beijing National Research Center for Condensed Matter Physics, and Institute of Physics, Chinese Academy of Sciences, Beijing 100190, China}
\author{Rui Yu}
 \email{yurui@whu.edu.cn}
\affiliation{School of Physics and Technology, Wuhan University, Wuhan 430072, China}
\author{Gang Xu}
 \email{gangxu@hust.edu.cn}
\affiliation{Wuhan National High Magnetic Field Center and School of Physics, Huazhong University of Science and Technology, Wuhan 430074, China}
\author{Kailun Yao}
\affiliation{Wuhan National High Magnetic Field Center and School of Physics, Huazhong University of Science and Technology, Wuhan 430074, China}

\date{\today}

\begin{abstract}
Analogues of the elementary particles,
Dirac fermions (DFs) in condensed matter have received extensive attention for both scientific interest and device applications.
In this work, we study the Dirac semimetals (DSMs) in the magnetic space groups (MSGs),
and find a new category of DSMs in the centro-symmetric type IV MSGs, where the Dirac points (DPs) are protected by the inversion symmetry, nonsymmorphic time-reversal symmetry and suitable rotation symmetry.
Moreover, we propose the interlayer antiferromagnetic (AFM) material EuCd$_2$As$_2$ as a promising candidate hosting only one pair of such DPs at Fermi level.
Many exotic topological states, such as the AFM triple point semimetal,
AFM topological insulator (TI) exhibiting the half-quantum Hall effect, can be derived from such AFM DSMs by breaking certain symmetry.
 Our results extend the range of DSM, and provide a platform to study the topological phase transition and the exotic properties of the AFM topological states.
\end{abstract}

\maketitle

\section{Introduction}
Massless DFs are one kind of the long-pursued elementary particles\cite{weyl1929elektron,volovik2003universe}.
While their existence remains elusive in particle physics, the realization of DFs in the DSMs\cite{murakami2007phase,kanedirac,wang2012dirac,wang2013three,vafek2014dirac,yang2014classification,young2017filling,bradlyn2017topological,po2017complete,watanabe2017structure}
has received extensive attention for both scientific interest and device applications\cite{koshino2010anomalous}. In the three-dimensional (3D) materials with both time-reversal symmetry $\mathcal{T}$ and inversion symmetry $ \mathcal{P}$,
each band energy is double degenerate.
If two double degenerate bands  linearly cross each other at discrete momentum point, such four-fold degenerate point is called Dirac point\cite{burkov2016topological},
whose low-energy excitation can be described by the massless relativistic Dirac equation.
Following such guideline, several 3D DSMs have been proposed and confirmed experimentally in nonmagnetic systems\cite{wang2012dirac,wang2013three,kanedirac,liu2014discovery,borisenko2014experimental,xiong2015evidence,novak2015large,steinberg}.

Generally, both time-reversal symmetry and inversion symmetry are necessary to protect such four-fold degenerate DPs in nonmagnetic materials.
Otherwise, the system will evolve into other exotic quantum states such as Weyl semimetals\cite{wan2011topological,xu2011chern,weng2015weyl,huang2015weyl,xu2015discovery,lv2015experimental,lv2015observation,wangwte2,yang2015weyl,sun2015topological,sun2015prediction,liang2016electronic, Nie03102017} or TIs\cite{zhang2011topological}.
However, an exception is proposed in a specific AFM configuration of CuMnAs recently\cite{tang2016dirac},
in which both $\mathcal{P}$ and $\mathcal{T}$ are broken but their combination $\mathcal{PT}$ is preserved.
So that Kramer's degeneracy is reserved  for the generic
momentum $k$, and DPs can exist in such kind of AFM system.
Thence, a natural question is whether there exists other type of DSMs in magnetic materials.

In the present paper, we find that the concept of DSM can be generalized to the centro-symmetric type IV MSGs
that break $\mathcal{T}$ but preserve $\mathcal{P}$ and nonsymmorphic time-reversal symmetry $\mathcal{T'} = \mathcal{T}\tau$,
where $\tau$ is a fractional translation operator that connects the black and white Bravais lattices.
Guided by this idea, a concrete example of such DSM, the interlayer AFM EuCd$_2$As$_2$, is predicted by the density functional theory (DFT) calculations.
Many exotic topological states can be derived from such AFM DSM.
When three-fold rotation symmetry ${C}_{3z}$ is broken, DSM phase can evolve into the AFM TI phase exhibiting the half-quantum Hall effect on the (001) surface as discussed by Joel Moore \emph{et al.}\cite{mong2010antiferromagnetic}.
In the case of breaking $\mathcal{P}$, AFM triple point semimetal, rather than Weyl semimetal, can be stabilized.
Our results extend the range of DSM, and provide one candidate to study the DSM and other exotic AFM topological states.

%\begin{figure}
%\includegraphics[clip,scale=0.110, angle=0]{fig1.pdf}
%\caption{Crystal structure of the interlayer AFM EuCd$_2$As$_2$. \textbf{(a)} The side view of the FM EuCd$_2$As$_2$. The blue, pink and light green balls represent Eu, Cd and As atoms, respectively. \textbf{(b)} The top view of the frustrated AFM EuCd$_2$As$_2$ with a $\sqrt{3}a\times\sqrt{3}a$ reconstruction. \textbf{(c)} The side view of the interlayer AFM EuCd$_2$As$_2$. The red arrows mean the directions of the magnetic momentum.
%}
%\end{figure}

\begin{figure}[h]
\includegraphics[clip,scale=0.110, angle=0]{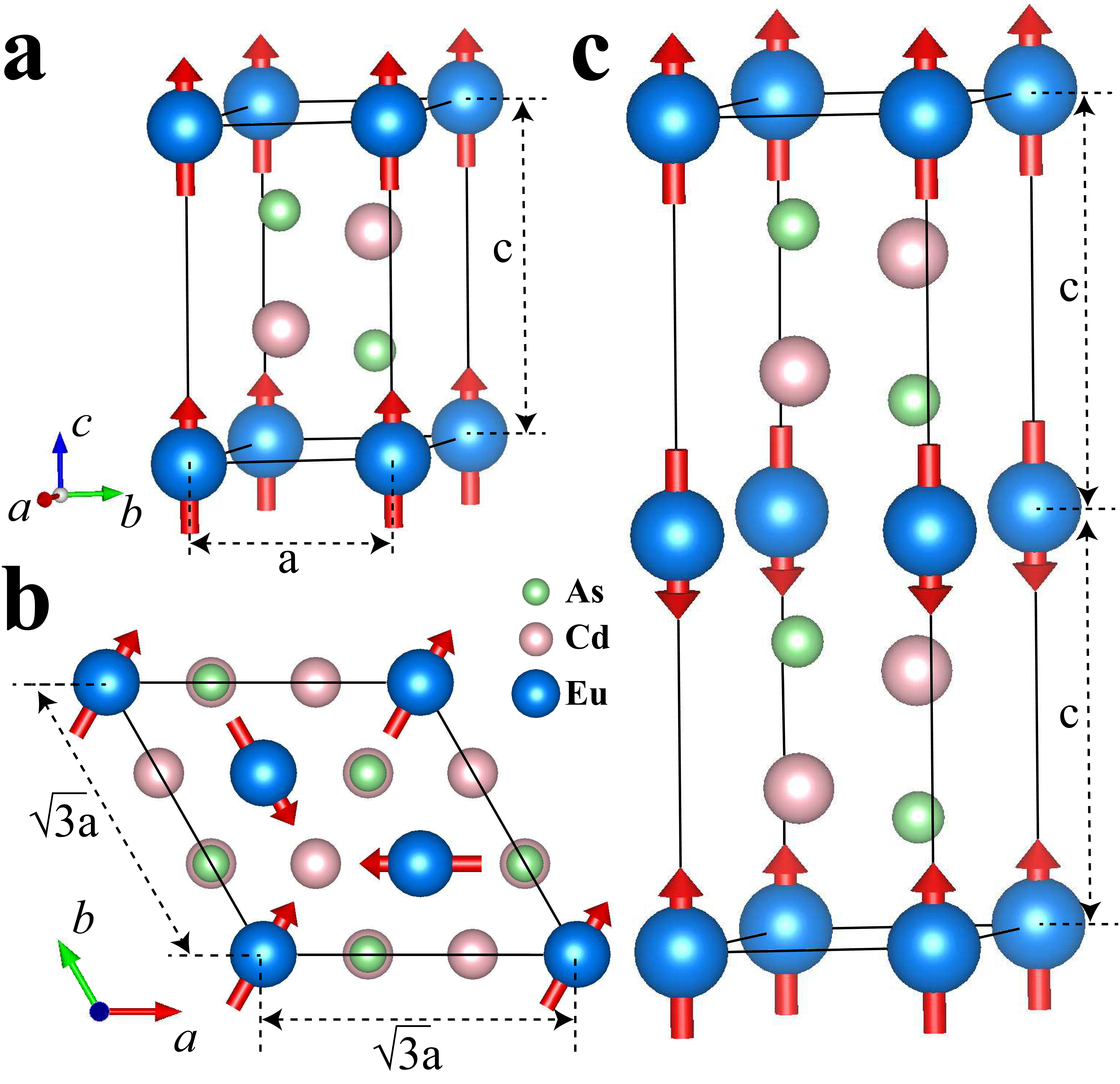}
\caption{Crystal structure of the interlayer AFM EuCd$_2$As$_2$. \textbf{(a)} The side view of the FM EuCd$_2$As$_2$. The blue, pink and light green balls represent Eu, Cd and As atoms, respectively. \textbf{(b)} The top view of the frustrated AFM EuCd$_2$As$_2$ with a $\sqrt{3}a\times\sqrt{3}a$ reconstruction. \textbf{(c)} The side view of the interlayer AFM EuCd$_2$As$_2$. The red arrows mean the directions of the magnetic momentum.
}
\label{structure}
\end{figure}

\section{COMPUTATIONAL METHODS}
The DFT calculations are carried out using the projector augmented wave method implemented in
Vienna $ab~ initio$ simulation package (VASP)\cite{kresse1996_1,kresse1996_2}.
Perdew-Burke-Ernzerhof type\cite{Perdew1996} of GGA + Hubbard U (GGA + U) approach\cite{liechtenstein1995density}
with U = 5 eV on Eu's 4\emph{f} orbitals is used to treat with the exchange and correlation potential.
SOC is taken into account self-consistently.
The cut-off energy of 500 eV and the 10 $\times$ 10 $\times$ 5 sampling of Brillouin Zone are used and carefully checked to ensure the convergence. All results are cross-checked with the full-potential linearized-augmented plane-wave method implemented in WIEN2k package.
MLWFs for the $p$ orbitals of As and $s$ orbitals of Cd have been generated and used to calculate the surface states iteratively\cite{sancho1984quick,sancho1985highly}.

\section{RESULTS}

\subsection{New type of DSM in type IV MSGs}
In order to search for new type of DSMs, we have checked all types of MSGs. The Kramer's degeneracy at generic momentum $k$ is an essential requirement for the realization of the DPs in solids. Therefore, any MSG that may host the DPs should contain time reversal operation $\mathcal{T}$ (or its combined operation), which is the premise condition of Kramer's degeneracy. According to Ref.~\cite{bradley2010mathematical}, there is no $\mathcal{T}$ presenting in type I MSGs, where MSGs $\mathcal{M}$ are defined as ordinary space group $\mathcal{G}$ ($\mathcal{M = G}$). So DSM phase can never exist in type I MSGs. In type II MSGs, $\mathcal{M}$ are defined as $\mathcal{G + TG}$. These MSGs are actually the space groups plus time reversal operation, in which the classification of all kinds of DSM phase have been discussed by Nagaosa \emph{et al}.\cite{yang2014classification}. In type III MSGs, $\mathcal{M}$ are defined as $\mathcal{H + T(G - H)}$, where $\mathcal{H}$ is a halving subgroup of the space group $\mathcal{G}$ and $\mathcal{(G - H)}$ contains no pure translations. In this case, the DPs can only exist in the type III MSGs that contain the combined operator $\mathcal{PT}$. AFM DSM CuMnAs discussed by Tang \emph{et al.}\cite{tang2016dirac} belongs to this situation. In type IV MSGs, $\mathcal{M}$ are defined as $\mathcal{G + T\{\mathbf{e}|\tau\}G}$, where $\mathbf{e}$ is the identity operation and $\mathcal{\tau}$ is the extra translation connecting the black (up spin) and white (down spin) sublattices. Obviously, there is no pure time reversal operation in type IV MSGs. The operation that changes the direction of the spin moment in the type IV MSGs becomes a nonsymmorphic form $\mathcal{T'} = \mathcal{T}\tau$. And we name $\mathcal{T'}$ as the  nonsymmorphic time-reversal symmetry to distinguish the ordinary time-reversal symmetry $\mathcal{T}$.

It is easy to identify that the representation of $\mathcal{T'}$ in the momentum space satisfies $\mathcal{T'}^2 = \mathcal{T}\tau\mathcal{T}\tau = -{\tau}^2=-e^{2ik\cdot \tau}$,
and itself can't ensure Kramer's degeneracy in the whole momentum space.
Fortunately, we find that the combination of $\mathcal{T'}$ and $\mathcal{P}$, \emph{i.e.}, $\mathcal{PT'}$,
whose square equals to -1 ($(\mathcal{PT'})^2 = \mathcal{PT}\tau\mathcal{PT}\tau = \mathcal{PT}\mathcal{P}(-\tau)\mathcal{T}\tau = \mathcal{P}^2\mathcal{T}^2 e^{-ik\cdot \tau} e^{ik\cdot \tau} = -1$),
is the sufficient condition to protect Kramer's degeneracy at the generic momentum $k$.
Therefore, we will focus on the centro-symmetric type IV MSGs that contain both $\mathcal{T'}$ and $\mathcal{P}$ in the following,
which may support DSM phase and have never been studied in the previous literatures.

In addition to the symmetry $\mathcal{T'}$ and $\mathcal{P}$, suitable crystal symmetry, such as rotation symmetry $C_n$,
is also needed to protect stable DPs in the centro-symmetric type IV MSGs.
Therefore, there are only two classes of locations that DPs can exist in the momentum space:
DPs on the rotation axis away from the time-reversal invariant momentum (TRIM) points (Class-I) and DPs on the TRIM points (Class-II).
In the presence of $\mathcal{P}$, $\mathcal{T'}$ and $C_n$ ($n=3$, 4 or 6), the classification of Class-I DSMs in the centro-symmetric type IV MSGs is homologous to the discussion given by Nagaosa \emph{et al.}\cite{yang2014classification},
because the little group of $k$ points on
the rotation momentum-axis is isomorphic with that in the type II MSGs,
both satisfying $C_n^n=-1$, $\mathcal{(PT')}^2 = -1$ and $[C_n, \mathcal{PT'}] = 0$.
As an illustrative example of the Class-I DSM in type IV MSGs,
the interlayer AFM EuCd$_2$As$_2$ will be addressed in this paper,
and its derivative exotic states, such as AFM TI and triple point semimetal will be discussed too.

\begin{figure}
\includegraphics[clip,scale=0.110, angle=0]{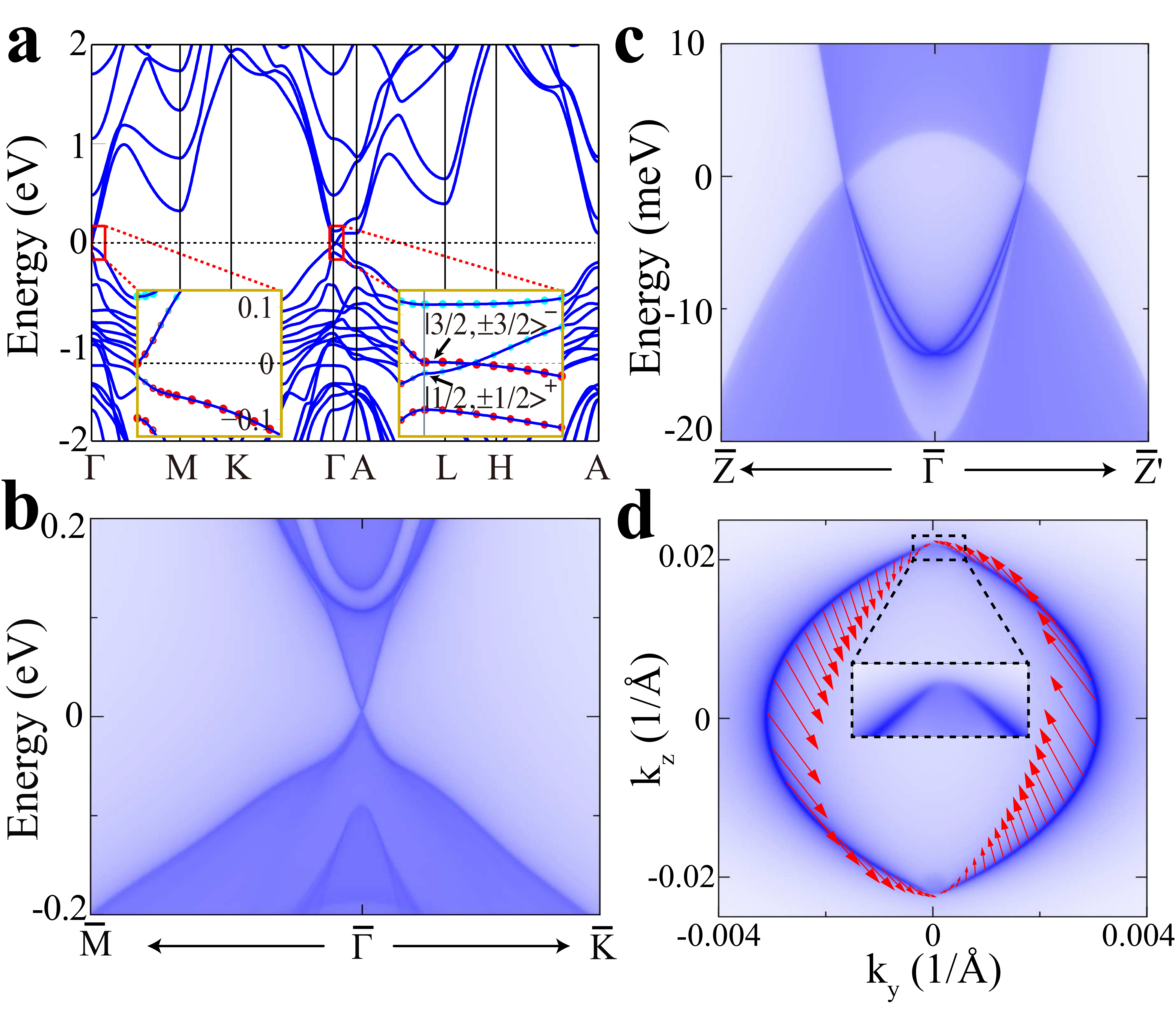}
\caption{The electronic structure in the interlayer AFM EuCd$_2$As$_2$. \textbf{(a)} The GGA+U+SOC calculated band structures of the interlayer AFM EuCd$_2$As$_2$. The insets are the zoom-in of the band structures around the $\Gamma$ point to show the DP obviously, where the red and the light blue dots represent the projections of the As $p$ and Cd $s$ orbitals, respectively. \textbf{(b)}, \textbf{(c)} are the calculated surface states of  the interlayer AFM EuCd$_2$As$_2$ on the (001) and (100) faces, respectively. \textbf{(d)} The Fermi arcs and their spin textures on the (100) face of the interlayer AFM EuCd$_2$As$_2$, where the inset reveals the discontinuity at the DPs between two Fermi arcs.
}
\label{fig2}
\end{figure}

\begin{figure}
\center
\includegraphics[clip,scale=0.102, angle=0]{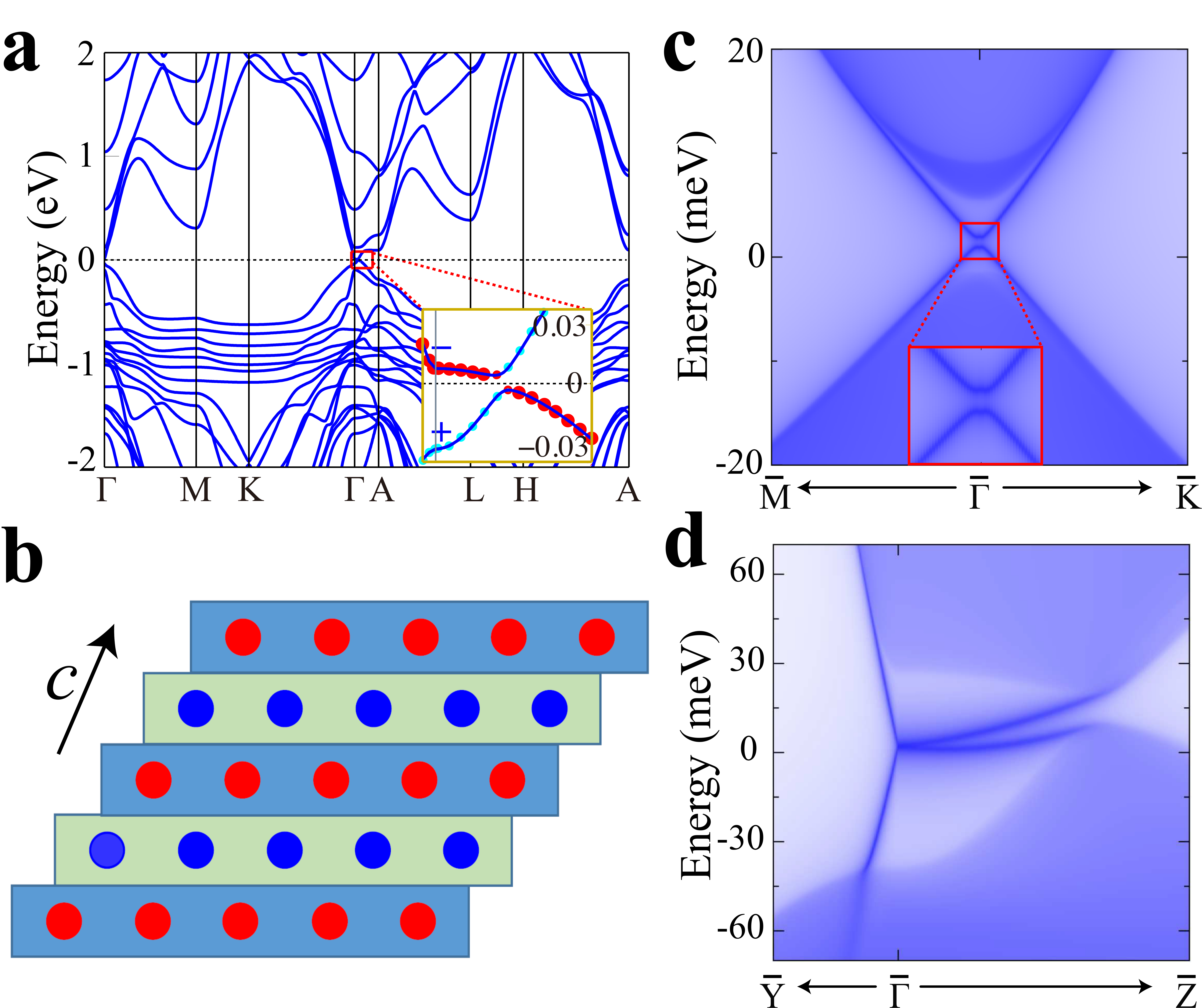}
 \caption{The derivative exotic topological states from the AFM DSM EuCd$_2$As$_2$ by breaking ${C}_{3z}$. \textbf{(a)} The GGA+U+SOC calculated band structures of the interlayer AFM EuCd$_2$As$_2$ in the case of ${C}_{3z}$ breaking. The insets are the zoom-in of band structures along the $\Gamma-A$ direction to exhibit the insulating gap. \textbf{(b)} Schematic of the AFM TI by stacking the 2D Chern insulators along the $c$-axis, where the red and blue balls represent the up spin layers (Chern number $C = 1$) and down spin layers ($C = -1$), respectively. \textbf{(c)}, \textbf{(d)} The calculated surface states of the AFM TI on the (001) and (100) faces, respectively. The inset of \textbf{c} shows the intrinsic gapped surface states clearly.
}
\label{fig3}
\end{figure}

Considering the representation ${\mathcal{T'}}^2=-e^{2ik\cdot \tau}$,
two cases of TRIM points need to be differentiated depending on the value of the $k\cdot\tau$:
Case1, TRIM points with $k\cdot\tau = 0$ or $\pi$; Case2, TRIM points with $k\cdot\tau = \pi/2$.
For Case1, the algebraic relations of $\mathcal{P}$ and $\mathcal{T'}$ are similar to the discussion in Ref.\cite{yang2014classification},
satisfying $\mathcal{P}^2 = 1$,  $\mathcal{T'}^2= -1$ and $[\mathcal{P}, \mathcal{T'}] = 0$.
When rotation symmetry $C_n$ ($n=2,4,6$) present,
single DP can be stabilized at the TRIM points of Case1,
because their little group in type IV MSGs is isomorphic with that in the type II MSGs,
\emph{i.e.}, $\{C_n,P\}=0$ and $[C_n,\mathcal{T}^\prime]=0$.
However, for the TRIM points with $k\cdot\tau=\pi/2$ (Case2),
the algebraic relations between $\mathcal{P}$ and $\mathcal{T'}$ are novel and particular,
$\mathcal{P}^2 = 1$,  $\mathcal{T'}^2= 1$, $\{\mathcal{P}, \mathcal{T'}\} = 0$ and $(\mathcal{P}\mathcal{T}^\prime)^2=-1$,
where the minus sign of $(\mathcal{P}\mathcal{T'})^2$ comes from the anti-commutative relation $\{\mathcal{P},\mathcal{T}^\prime\}=0$.
To the best of our knowledge, such kind of algebraic relations have never been addressed in the fermion system.
So the conclusion in Ref.~\cite{yang2014classification} is not applicable now.
In this paper, we just provide a proof that these algebraic relations prohibit all kinds of DPs with odd-order dispersions in type IV MSGs.
The general discussions of their topological character are left in the future work.

Let us begin by assuming a four-fold band degeneracy located at the Case2 TRIM points.
Without loss of the generality, one can write the four-dimensional $\mathcal{T'}=\tau_0\sigma_x K$, $\mathcal{P}=\tau_0\sigma_z$,  $\mathcal{PT'}=i\tau_0\sigma_y K$,
where $\tau_0$ indicates the two-dimensional identity matrix for the orbital basis,
$\sigma_{x,y,z}$ are the Pauli matrices describing spin degrees of freedom, and $K$ is complex conjugation operator.
It is obvious that, besides the identity matrix $\tau_0\sigma_0$, only the following five $\gamma$-matrices,
\emph{i.e.}, $\sigma_0\tau_x$, $\sigma_0\tau_z$, $\sigma_x\tau_y$, $\sigma_z\tau_y$, $\sigma_y\tau_y$
can be used to construct the Hamiltonian yielding to the requirement $[H(k), \mathcal{PT'}] = 0$, where $H(k)=\sum_{i=1,5}f_i(k) \gamma_i$, and $f_i(k)$ are real functions of $k$.
Furthermore, due to the requirement of  $\mathcal{P}H(k)\mathcal{P}^{-1}=H(-k)$,
only two $\gamma$-matrices, $\sigma_x\tau_y$ and $\sigma_y\tau_y$, which are anti-commute with $\mathcal{P}=\tau_0\sigma_z$, could couple to the odd type function $f(k)$.
On the other hand, to get a linear DP, we need three such kind of $\gamma$-matrices.
To the leading order, the effective Hamiltonian around the DP can be written as $H=\sum_{i,j}^{3}k_iA_{ij}\gamma_j$
and the dispersion is given by $E(k)=\pm\sqrt{\sum_j^3(\sum_i^3 k_i A_{ij})^2}$.
Here $A$ is a $3\times3$ matrix with full rank.
If we have only two (or one) such $\gamma-$matrices, or, equivalently, $A$ is  a rank-2   (rank-1) matrix, then the velocity of dispersion must vanish along one (two) particular direction,
which can be indicated by the zero eigenvector(s) of $A$.
Similarly, it can be proved that all kinds of odd-order DPs can not exist at the Case2 TRIM points in type IV MSGs,
and this conclusion does not depend on the representation choice of $\mathcal{T'}$ and $\mathcal{P}$.

\subsection{AFM DSM EuCd$_2$As$_2$}
Guided by the DSM classification in the type IV MSGs, we discover that a pair of Class-I DPs (belonging to type IV MSGs)
can be hosted at the Fermi level in the interlayer AFM EuCd$_2$As$_2$.
As shown in Fig. 1a, EuCd$_2$As$_2$ crystallizes into the CaAl$_2$Si$_2$-type structure with space group $D_{3d}^3$ ($P\bar{3}m1$)~\cite{artmann1996am2x2,goryunov2012esr,schellenberg2010121sb},
in which Cd$_2$As$_2$ layers are separated by the trigonal Eu layers.
Considering that Eu$^{2+}$ has a half-filled $4f$-shell,
we have calculated three different magnetic configurations for EuCd$_2$As$_2$ by the generalized gradient approximation (GGA) + Hubbard
U (GGA+U) method with U = 5 eV,
including the ferromagnetic (FM), frustrated AFM and interlayer AFM configurations as shown in Fig. 1a, 1b and 1c, respectively.
All the magnetic moments proposed here are assumed along c-direction, which is a little contrast to recent observations \cite{rahn2018coupling}. This is because EuCd$_2$As$_2$ is a very soft magnet \cite{rahn2018coupling}. Our calculations indicate that it is easy to change its moment direction by enlarging the in-plane lattice constant. More details and discussions of the magnetic anisotropy are presented in the Appendix A.
The calculated total energies and moments are summarized in Table I. Our calculations demonstrate that all magnetic states are lower than the
nonmagnetic state about 6.3 eV/f.u., and the interlayer AFM is the most stable one, further lowering the total energy about 2 meV/f.u. than
the ferromagnetic states, which are consistent with recent experiment measurement very well\cite{wang2016anisotropic}.

 \begin{table*}[tp]
\caption{Total energies (in unit of eV/f.u.) of four different magnetic structures for EuCd$_2$As$_2$ calculated by GGA+SOC+U.
 The converged magnetic moments (in unit of $\mu_B$) of each Eu atom are given too.}
  \begin{tabular}{c c c c c c c}
\hline
        Config.             &~~ $\text{Eu}_1$    &~~$\text{Eu}_2$  &~~$\text{Eu}_3$  &~~Energy \\
\hline
        Nonmagnetic         &~~(0, 0, 0)         &~~(0, 0, 0)    	&~~(0, 0, 0)      &~~-18.021
 \\
        FM                    &~~(0, 0, 6.88)      &~~(0, 0, 6.88)	&~~(0, 0, 6.88)   &~~-24.368
\\
        Interlayer AFM      &~~(0, 0, 6.88)      &~~(0, 0, -6.88)	&~~               &~~-24.370
\\
        Frustrated AFM      &~~(3.44, 5.95, 0)    &~~(-6.87, 0, 0)	&~~(3.44, -5.95, 0)&~~-24.365
\\
\hline
\end{tabular}
\label{totenergy}
\end{table*}

The projected band structures of the interlayer AFM EuCd$_2$As$_2$ are shown in Fig. \ref{fig2}a.
Our DFT calculations indicate that the low-energy bands near the Fermi level are mainly contributed from the $p$ orbitals of As atoms and the $s$ orbitals of the Cd atoms.
In particular, the double degenerate $s$-$s$ bonding states of Cd atoms (even parity) invert with the $p$-$p$ antibonding states of As atoms (odd parity) at the Fermi level around the $\Gamma$ point.
For the first glance, it is a little surprising that the crossing along $\Gamma-A$ line has no gap but a stable four-fold degenerate DP, because $\mathcal{T}$ is broken.
However, after a detailed symmetry analysis, one can find a nonsymmorphic time-reversal symmetry $\mathcal{T'} = \mathcal{T}\oplus c$,
connecting the up spin momentum layer at $z = 0$ and the down spin momentum layer at $z = c$, exists in this interlayer AFM system.
The MSGs of the interlayer AFM EuCd$_2$As$_2$ can be expressed as $D_{3d}^4 \oplus \mathcal{T'}D_{3d}^4$,
whose generators include $\mathcal{T'}$, $\mathcal{P}$, $C_{3z}$ and two-fold screw $C'_{2x}=C_{2x} \oplus c$.
Combining $\mathcal{T'} = \mathcal{T} \oplus c$ with $\mathcal{P}$ together,
the antiunitarity of $\mathcal{PT'}$ would prohibit the hopping terms between the nonsymmorphic time-reversal pair of states,
such as $|3/2,\pm3/2\rangle$ or $|1/2,\pm1/2\rangle$. So that every energy state is double  degenerate in such interlayer AFM system.
As we discussed above, it is possible to host the DPs with the help of proper rotation symmetry,
which is exactly what happens on the $\Gamma-A$ line in the interlayer AFM EuCd$_2$As$_2$.

The little group of $k$ points on the $\Gamma-A$ line can be described as $C_{3v} \oplus \mathcal{PT'}C_{3v}$. When
spin-orbit coupling (SOC) is included, the topology of the system and the band inversion are dominated by the four states
$|3/2,\pm3/2\rangle^{-}$ from the $p$-$p$ antibonding states of As and $|1/2,\pm1/2\rangle^{+}$ from the $s$-$s$ bonding states of Cd.
Under the symmetry restrictions, an effective $4 \times 4$ $k\cdot p$ model
(in the order of $|1/2,1/2\rangle^{+}$, $|3/2,3/2\rangle^{-}$, $|1/2,-1/2\rangle^{+}$, $|3/2,-3/2\rangle^{-}$)
around the $\Gamma$ point can be written in the following form,
\begin{equation*}
  \displaystyle
  \begin{aligned}
   H =\epsilon_{0}(k)+
   \left[
      \begin{array}{cccc}
              M(k)     &    A k_{+}  &    0       &    B k_{+}    \\
             Ak_{-}  &      -M(k)      &  B k_{+}   &      0        \\
               0     &   B k_{-}   &   M(k)       &  - A k_{-}    \\
           B k_{-}   &        0    &  -A k_{+}  &      -M(k)   \\
      \end{array}
    \right]
  \end{aligned}
\end{equation*}
where $\epsilon_{0}(k) = C_{0}+C_{1}k^{2}_{z}+C_{2}(k^{2}_{x}+k^{2}_{y})$, $k_{\pm} = k_{x}\pm ik_{y}$
and $M(k)  = M_{0}-M_{1}k^{2}_{z}-M_{2}(k^{2}_{x}+k^{2}_{y})$ with parameters $M_0$, $ M_1$, $M_2 < 0$ to guarantee the band inversion.
We note that our effective model is very similar to the Hamiltonian in Na$_3$Bi,
except that the off-diagonal terms take the leading order $B k_{\pm}$ rather than high-order $Bk_zk_{\pm}^2$ as in Na$_3$Bi. By diagonalizing the above Hamiltonian, we can get two double degenerate eigenvalues
$E_{\pm} =\epsilon_{0} \pm \Delta$ with $\Delta =\sqrt{(A^{2}+B^{2})(k_{x}^{2}+k_{y}^{2})+M(k)^{2}}$.
Evaluating the eigenvalues, it is clear that two linear DPs at $k^c = (0, 0, \pm \sqrt{M_0/M_1})$ exist on the $\Gamma-A$ line, which are consistent with our DFT calculations.
 This fact is confirmed by the calculated surface states (Fig. \ref{fig2}b, \ref{fig2}c) and Fermi arcs (Fig. \ref{fig2}d)
based on the Green's functions of the semi-infinite system, which are constructed by the maximally localized Wannier functions.
The (001) surface states plotted in Fig. \ref{fig2}b exhibit a clear band touching at the $\bar{\Gamma}$ point and Fermi level,
where a pair of DPs are projected to the same point on the (001) face.
More evidences supporting the DSM phase come from the Fermi arcs on (010) face as plotted in Fig. \ref{fig2}c and Fig. \ref{fig2}d,
where a pair of arc states connect the DPs from the bulk state unambiguously.
As shown in the inset of Fig. \ref{fig2}d, even though the Fermi arcs look like closed,
their Fermi velocities are discontinuous at the DPs.
Such kind of closed Fermi arcs in DSMs have been discussed by Zhijun Wang \emph{et al.} \cite{wang2012dirac} and Peizhe Tang \emph{et al.}\cite{tang2016dirac}.

\begin{figure}
\center
\includegraphics[clip,scale=0.105, angle=0]{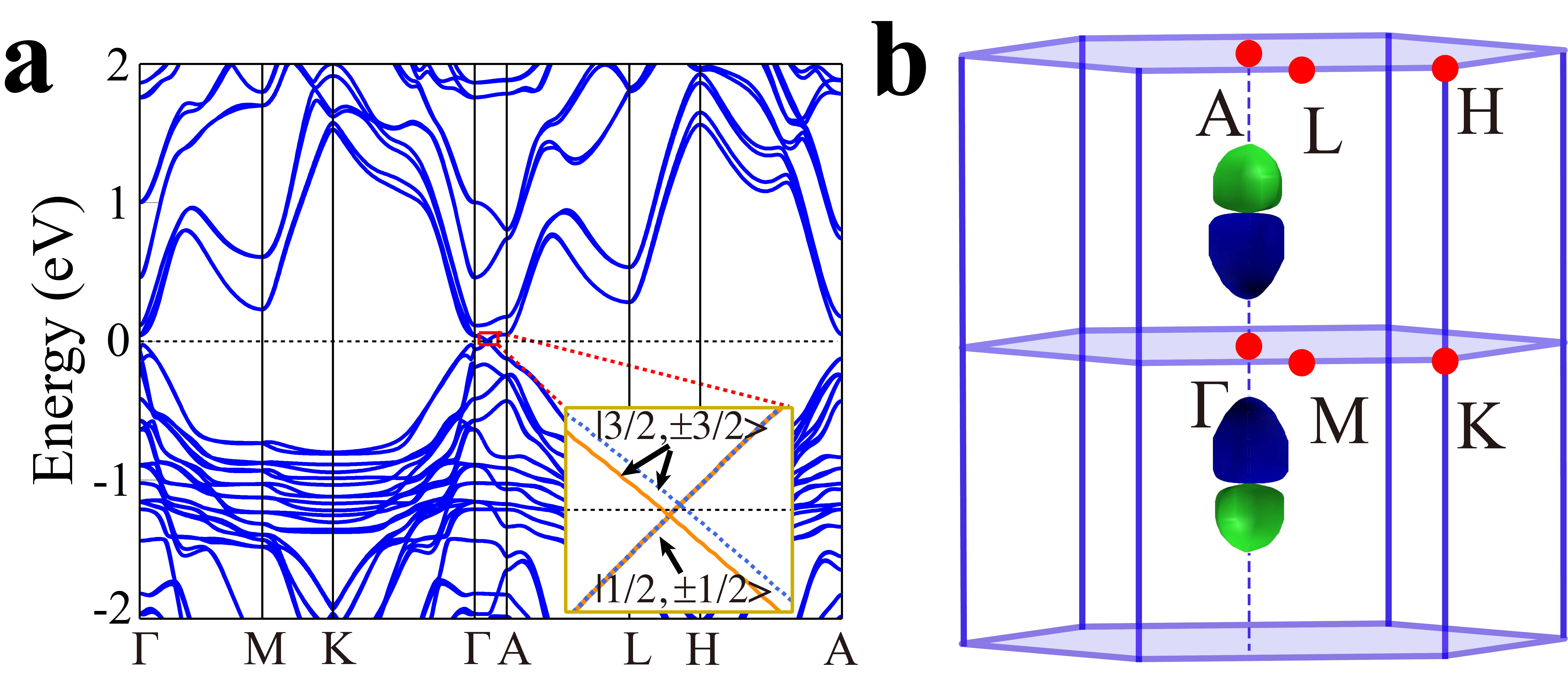}
\caption{The derivative exotic topological state from the AFM  DSM EuCd$_2$As$_2$ by breaking $\mathcal{P}$. \textbf{(a)} The GGA+U+SOC calculated band structures of the interlayer AFM EuCd$_2$As$_2$ in the case of $\mathcal{P}$ breaking (triple point semimetal). The insets are the zoom-in of band structures along the $\Gamma-A$ direction to exhibit the triple points distinctly. \textbf{(b)} The calculated Fermi surfaces of triple point semimetal derived from the interlayer AFM EuCd$_2$As$_2$ with $\mathcal{P}$ breaking, where the Fermi surfaces are magnified 60 times to exhibit them and their tangency visibly.
}
\label{fig4}
\end{figure}

\subsection{TI phase and triple point semimetal in AFM system}
In addition to the generic features discussed above, our AFM DSM has its own uniqueness. Such
uniqueness can be reflected by its derivatives, which makes our AFM DSM different from
Na$_3$Bi and CuMnAs.

We would like to discuss the derivative TI phase from the AFM DSM EuCd$_2$As$_2$ first.
As discussed above,
$C_{3z}$ is important for the stability of the DPs in the interlayer AFM EuCd$_2$As$_2$.
When this symmetry is broken, $j_z$ is no longer a good quantum number on $\Gamma-A$ line.
So that the hopping terms between $|j_z = \pm1/2\rangle$ and $|j_z = \pm3/2\rangle$ can be introduced,
and the system evolves into the strong TI phase due to the inverted band structure.
One can achieve this object by applying the uniaxial strain or by tuning the magnetic moment to the in-plane direction, as shown in Appendix B\cite{rahn2018coupling}.

In Fig. \ref{fig3}a, we plot the band structures by enlarging the $a$-axis 1\%,
where an insulating gap of 9 meV opens up clearly.
The TI phase realized in Fig. \ref{fig3}a has unique topological properties.
It is not the conventional 3D TI as Bi$_2$Se$_3$,
but the AFM TI protected by $\mathcal{T'}$ as discussed by Joel Moore \emph{et al.}\cite{mong2010antiferromagnetic}.
To illustrate the difference, one can see the (001) surface states plotted in Fig. \ref{fig3}c,
where the surface states are intrinsically gapped at the $\bar{\Gamma}$ point,
because $\mathcal{T'}$ is broken when open boundary is applied on the $c$-axis.
However, when open boundary is applied to the other directions, such as $a$-axis, where $\mathcal{T'}$ is preserved,
the surface states remain gapless as that in the conventional TI (see Fig. \ref{fig3}d).
These characters conform with the discussion of the AFM TI~\cite{mong2010antiferromagnetic} exactly,
which can be taken as a product state of the two-dimensional (2D) Chern insulators stacked along the $c$-axis,
and each pair of the nearest neighbors are connected by $\mathcal{T}' = \mathcal{T} \oplus c$ (Fig. \ref{fig3}b).
Therefore, a nontrivial AFM $\mathbb{Z}_2$ invariant related to $\mathcal{T'}$ can be defined,
and the half-quantum Hall effect can be realized on the intrinsically gapped (001) face of such AFM TI~\cite{mong2010antiferromagnetic} (Fig. \ref{fig3}c).

The other big difference between our AFM DSM and the other DSMs (Na$_3$Bi and CuMnAs) can be reflected by breaking $\mathcal{P}$. As we all know, the DPs usually split into two pairs of Weyl points with opposite chirality in the conventional DSM when $\mathcal{P}$ is broken.
However, two pairs of triple points protected by the $C_{3v}$ little group (consisting of $C_{3z}$ and $M_x$) are obtained in our AFM DSM when $\mathcal{P}$ is broken.
Such results are plotted in Fig. \ref{fig4}a, in which $|j_z = \pm3/2\rangle$ states are split, while the $|j_z = \pm1/2\rangle$ states remain double degenerate.
The origin of the triple point semimetal realized in the $\mathcal{P}$ breaking EuCd$_2$As$_2$ can be understand as following.
In absence of $\mathcal{PT^\prime}$ symmetry, the little group of $k$ points on
$\Gamma-A$ reduces to the
magnetic point group $C_{3v}$, which has one 2D irreducible representation $E_{1/2}$ ($|\pm 1/2\rangle$)
and two one-dimensional irreducible representations $E_{3/2}$ ($\frac{1}{\sqrt{2}} |3/2\rangle \pm \frac{i}{\sqrt{2}}|-3/2\rangle$).
Therefore, the degeneracy between $|\pm 3/2\rangle$ states,
which is originally protected by $\mathcal{PT^\prime}$, is broken;
while the degeneracy between $|\pm 1/2\rangle$ remains, leading to two pairs of triple points on the $\Gamma-A$ line naturally.
We calculate the Fermi surfaces of the triple point semimetal phase and plot them in Fig. \ref{fig4}b,
where two pairs of tangent Fermi pockets exist, and each pocket encloses one triple point.
Similar to the nonmagnetic triple point semimetal\cite{zhu2016triple,weng2016coexistence,weng2016topologicalTaN},
two touching Fermi pockets hold opposite spin winding number.
Finally, it's worthy to note that this is the first reported triple point semimetal in the magnetic material. The topological property of such magnetic topological material is usually related to the specific magnetic order, which provides a new mechanism to tune the topological phase transition by changing the magnetic order or the external magnetic field.

\section{CONCLUSIONS}
In summary, we generalized the concept of DSM to the centro-symmetric type IV MSGs,
where the antiunitarity of the product between $\mathcal{P}$ and $\mathcal{T'}$, \emph{i.e.} $(\mathcal{PT'})^2 = -1$ is essential for  Kramer's degeneracy and the AFM DPs.
Based on DFT calculations, we propose that the interlayer AFM EuCd$_2$As$_2$ is a candidate of such AFM DSM.
Many exotic topological states can be derived from the AFM DSMs.
For example, when the three-fold rotation symmetry is broken, it can evolve into the AFM TI discussed by Joel Moore \emph{et al.}\cite{mong2010antiferromagnetic},
where the half-quantum Hall effect can be realized on the intrinsically gapped (001) face.
If $\mathcal{P}$ is broken, it can result in the triple point semimetal phase, rather than the Weyl semimetal.
Our results provide a new direction to study the DSM and other exotic AFM topological states.

~\\
  \indent We thank Tian Qian and Yulin Chen for useful discussions. This work is supported by the Ministry of Science and Technology of China (2018YFA0307000);
  G. X. and R. Y. are supported by the National Thousand-Young-Talents Program
 and the National Natural Science Foundation of China; S. N. is supported by Stanford Energy 3.0.

~\\
\emph{Note added.}
The evidence of a Dirac-cone type dispersion in EuCd$_2$As$_2$ has been observed recently in the angle-resolved photoemission spectroscopy (ARPES) measurements by Ma  \emph{et al.}\cite{ma2018}, which greatly support our theoretical predictions.

\appendix

\section{Magnetic anisotropic energy of EuCd$_2$As$_2$}
\begin{figure}
\includegraphics[clip,scale=0.30, angle=0]{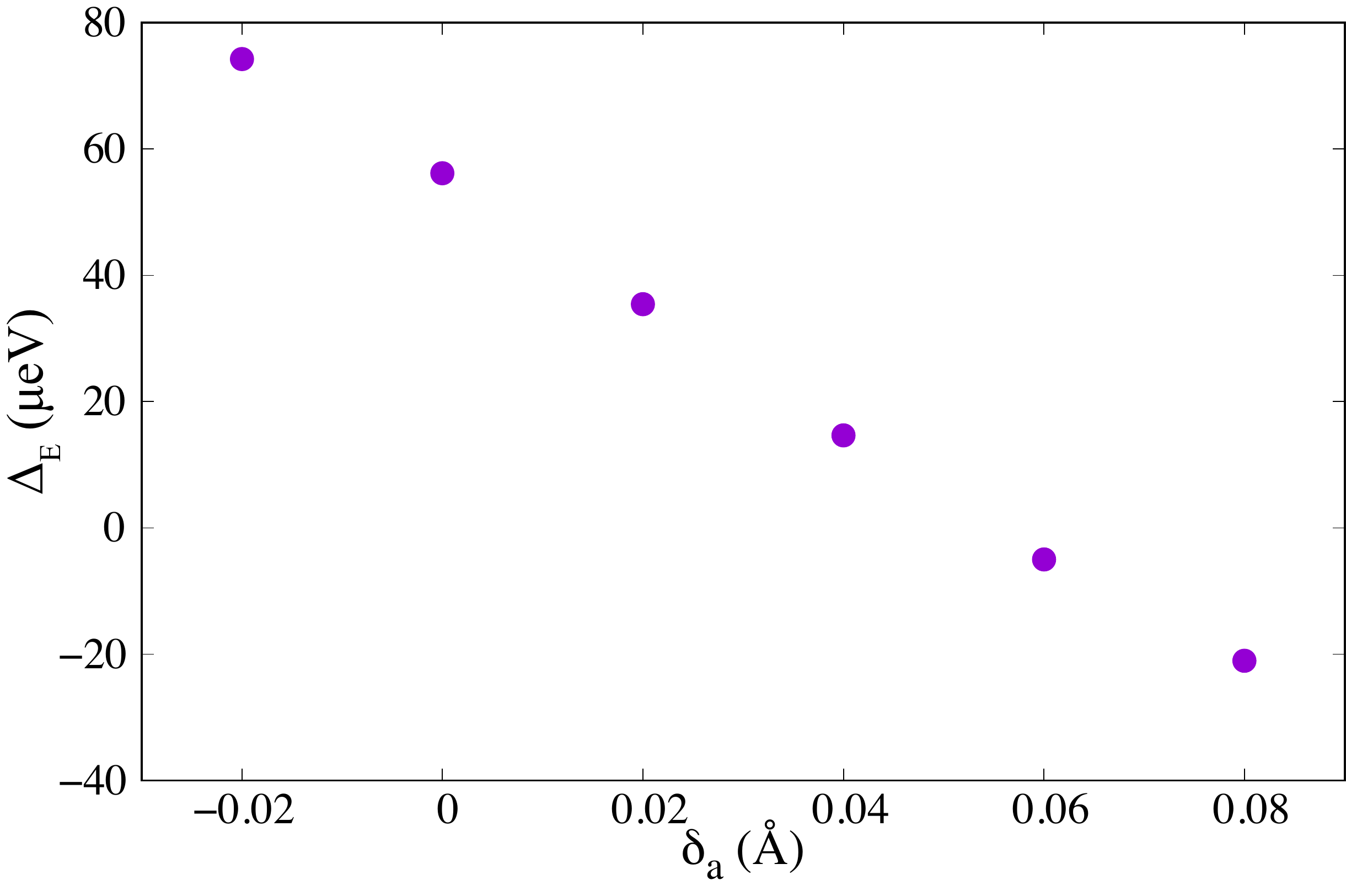}
\caption{The anisotropic energy $\Delta _E$ versus the in-plane lattice constant changing $\delta_a$. $\Delta _E$ is defined as the total energy difference between the out-of-plane A-AFM and the in-plane A-AFM state, i.e., $\Delta _E=E(AFM_{out-of-plane})-E(AFM_{in-plane})$, and
$\delta_a$ is defined as  $a-a_0$, where $a_0$=4.4 \AA~is the experimental lattice constant
}
\label{figs1}
\end{figure}

Recently, a related experimental work \cite{rahn2018coupling} claimed that the ground state of
EuCd$_2$As$_2$ is an A-type AFM with in-plane magnetic moments, which is a little contrast to
our results. However, they also found that EuCd$_2$As$_2$ is a very soft magnetic material,
indicating that the energy difference of magnetic anisotropy is very small. Therefore, it should be
easy to tune the direction of the magnetic moment by modulating the external environment.
Based on DFT calculations, we found that the ground state will
change from the in-plane A-AFM  into the out-of-plane A-AFM  by slightly stretching the in-plane
lattice constant $a$, as shown in Fig. \ref{figs1}. It shows that the total energy
difference ($\Delta_E$) between these two AFM configurations is less than 1 meV, which is
consistent with the experimental results very well. Moreover, when the in-plane lattice
constant is increased by 0.06 \AA, the direction of the magnetic moment will change from the
in-plane A-AFM to the out-of-plane A-AFM configuration. In experiment, one can grow the
EuCd$_2$As$_2$ sample on a suitable large lattice constant substrate to get the out-of-plane
A-AFM state.

\begin{figure}
\includegraphics[clip,scale=0.50, angle=0]{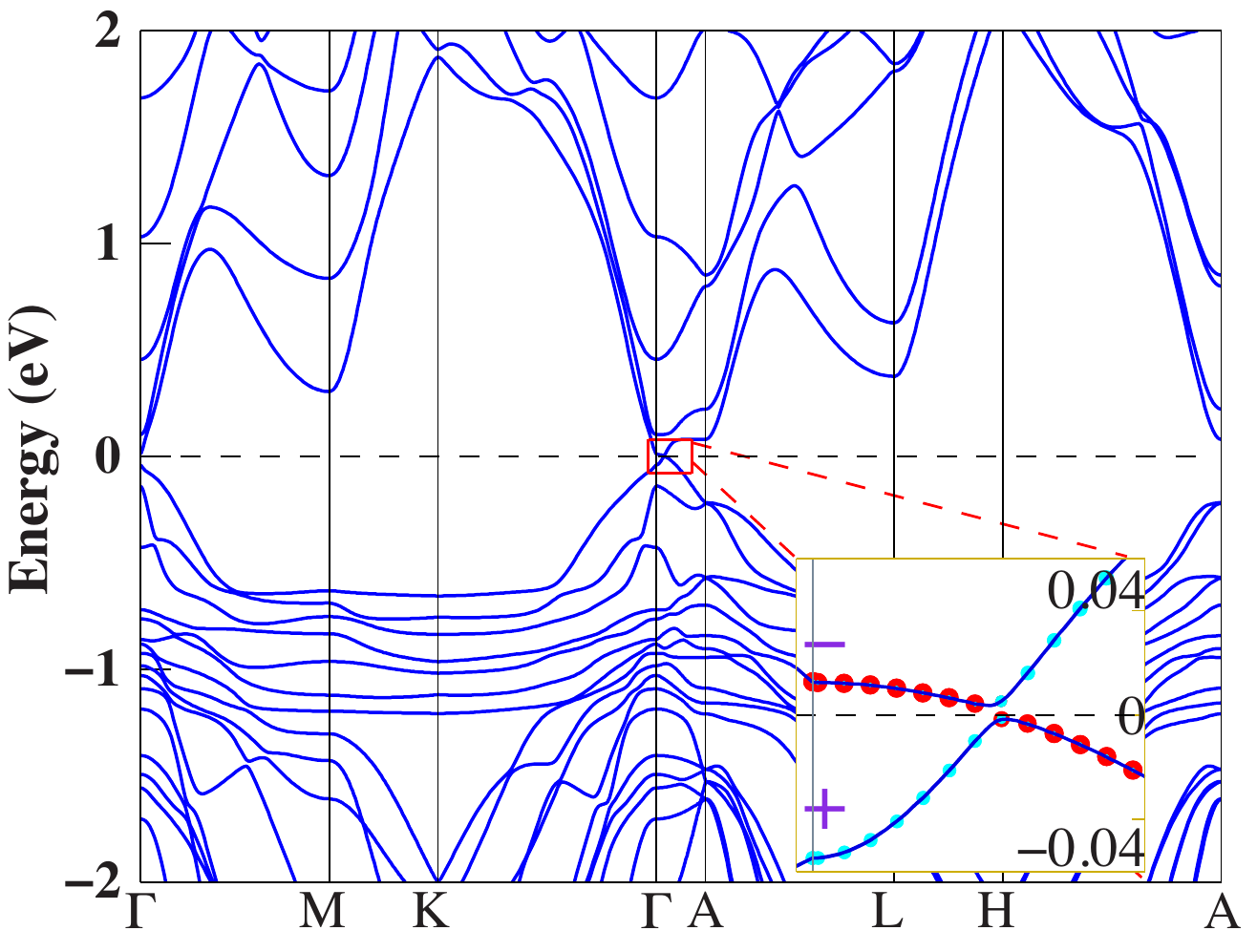}
\caption{The band structures of EuCd$_2$As$_2$ with in-plane A-AFM structure. The inset shows the band gap opening at the Dirac point.
}
\label{figs2}
\end{figure}

\section{AFM TI for EuCd$_2$As$_2$ with in-plane A-AFM}
It is worthy to note that EuCd$_2$As$_2$ is also very interesting even the ground state
is stabilized in the in-plane A-AFM state. Due to the fact of the existence of the band
inversion and the gap opening, the in-plane A-AFM EuCd$_2$As$_2$ is actually a 3D AFM
topological insulator, which is just the same state as we showed in the Fig. 3 of the main
text. The calculated band structures based on the in-plane A-AFM configuration is shown in
Fig. \ref{figs2}, where the red and the light blue dots represent the projections of the
As $p$ and Cd $s$ orbitals, respectively. It is clear that the band inversion still survive
and the band gap opening happens at the Dirac point (with band gap about 4 meV), which are
consistent with previous results \cite{rahn2018coupling}. After the band gap opening,
EuCd$_2$As$_2$ changes from the AFM Dirac semimetal to the AFM topological insulator.
This means that one can modulate the topological phase transition in the AFM topological
matters by modifying the direction of the magnetic moment, which provides us a new
mechanism to adjust the topological properties.

\end{document}